\documentclass[superscriptaddress,twocolumn,preprintnumbers,amsmath,amssymb,aps]{revtex4}

\pdfoutput=1
\usepackage[utf8]{inputenc}
\usepackage[USenglish]{babel}

\usepackage{amsmath,amssymb,amsthm,amsfonts}
\usepackage{MnSymbol}
\usepackage{mathrsfs}
\usepackage{slashed}
\usepackage{graphicx}
\usepackage{subfigure}
\usepackage{color,rotating}
\usepackage{dsfont}
\usepackage{setspace}
\usepackage{verbatim}
\usepackage{hyperref,doi}
\usepackage[export]{adjustbox}
\usepackage[normalem]{ulem}
\usepackage{natbib}
\usepackage{bbold}

\makeatletter
\def\l@subsubsection#1#2{}
\makeatother

\newcommand{\beq}{\begin{equation}}
\newcommand{\eeq}{\end{equation}}
\newcommand{\beqa}{\begin{eqnarray}}
\newcommand{\eeqa}{\end{eqnarray}}
\newcommand{\bfc}{\begin{figure}[t]\begin{center}}
\newcommand{\efc}{\end{center}\end{figure}}

\def\fig#1{Fig.~\ref{#1}}

\def\eq#1{(\ref{#1})}
\def\sec#1{section~\ref{#1}}
\def\0#1#2{\frac{#1}{#2}}  %% fractions

\graphicspath{{./figures/}}
\newcommand{\tr}{\mathrm{tr}}

\newcommand{\be}{\begin{eqnarray}}
\newcommand{\ee}{\end{eqnarray}}

\begin{document}

\title{Machine Learning of Explicit Order Parameters:\\ From the Ising Model to SU(2) Lattice Gauge Theory}

\author{Sebastian J. Wetzel} \affiliation{Institut f\"ur Theoretische
  Physik, Universit\"at Heidelberg, Philosophenweg 16, 69120
  Heidelberg, Germany}
\author{Manuel Scherzer} \affiliation{Institut f\"ur Theoretische
  Physik, Universit\"at Heidelberg, Philosophenweg 16, 69120
  Heidelberg, Germany}
%%%%%%%%%%%%%%%%%%%%%%%%%%%%%%%%

\begin{abstract}

We present a procedure for reconstructing the decision function of an artificial neural network as a simple function of the input, provided the decision function is sufficiently symmetric.
In this case one can easily deduce the quantity by which the neural network classifies the input.
The procedure is embedded into a pipeline of machine learning algorithms able to detect the existence of different phases of matter, to determine the position of phase transitions and to find explicit expressions of the physical quantities by which the algorithm distinguishes between phases.
We assume no prior knowledge about the Hamiltonian or the order parameters except Monte Carlo-sampled configurations.
The method is applied to the Ising Model and SU(2) lattice gauge theory.
In both systems we deduce the explicit expressions of the known order parameters from the decision functions of the neural networks.

\end{abstract}

\maketitle

%%%%%%%%%%%%%%%%%%%%%%%%%%%%%%%%

\section{Introduction}
\noindent 

Machine learning enables computers to learn from experience and generalize their gained knowledge to previously unseen problems. The development of better hardware and algorithms, most notably artificial neural networks, propelled machine learning to one of the most transformative disciplines of this century. Nowadays such algorithms are used to classify images \cite{Krizhevsky2012}, to recognize language \cite{Hinton2012} or to beat humans in complex games \cite{Silver2016}. Recently, machine learning has even been successfully employed to tackle highly complex problems in physics \cite{Carleo2017,Torlai2016,Deng2016,Crawford2016,Deng2017,Gao2017,Torlai2017,Aoki2016,Kasieczka2017,Huang2017,Liu2017,Portman2016}.
It is now possible to classify phases of matter in the context of supervised learning \cite{Carrasquilla2017,Broecker2016,Li2016,Chng2016,Zhang2016,Schindler2017} only from Monte Carlo samples. Phases can also be found without any information about their existence by unsupervised learning \cite{Wang2016,Wetzel2017,Nieuwenburg2017,Hu2017}. It is a difficult task to interpret what machine learning algorithms learn to classify phases, although a first progress was made using support vector machines \cite{Ponte2017}. The problem is not inherent to physics, since machine learning algorithms, especially artificial neural networks, exploded in complexity and application, and thus the difficulty of interpreting their decisions increased rapidly. There is still no comprehensive theoretical understanding of what is learned by them \cite{Towell1992,Mahendran2014,Ghahramani2015,Shwartz-Ziv2017}. 

In the context of physical phase transitions, we open the neural network black-box, and show that the learned decision functions originate from physical quantities. Not only can we interpret these physical quantities, we can explicitly deduce them without prior knowledge.

To this end we propose a new type of neural network, called correlation probing neural network. It can reduce the complexity of the decision function if it is sufficiently symmetric. Physical quantities are typically highly symmetric. Therefore, this network is ideal for probing whether a physical quantity is responsible for the learned decision function. After reducing the complexity with the correlation probing network, we show that it is possible to fully reconstruct the explicit mathematical expression of the decision function. From this expression one can then easily extract the quantities by which the neural network distinguishes between phases. This procedure is introduced at the Ising Model, where we find that the neural network predicts the phase by the magnetization and the expected energy per spin. We then demonstrate the power of our method at SU(2) lattice gauge theory, which is a QCD-like theory showing confinement. In this case we find that the decision function is based on a non-local and non-linear order parameter, the so-called Polyakov loop.

Our method might find application where phase transitions are inaccessible by other tools. Since machine learning methods can detect phase transitions in regions of phase diagrams that show a sign problem \cite{Broecker2016}, a possible application in QCD could be the examination of the chiral phase transition or the color-superconducting phase \cite{Alford2008,Stephanov2005,Kogut2004}. Our method could be employed to study the pseudogap in the two dimensional Hubbard model \cite{Varma2006,Taillefer2010,Wang2002,Huefner2008}. Another application could be the examination of multicritical points, for which the Hubbard model provides a prominent example in the regime of competing d-wave and antiferromagnetic order \cite{Friederich2011,Maier2005,Raghu2011,Halboth2000}.

%%%%%%%%%%%%%%%%%%%%%%%%%%%%%%%%%%%%%%%%%%%%%%
\section{Models}
\subsection{Ising Model}
\label{Sec:Ising}
\noindent The Ising model was originally formulated as a model for interactions between magnetic dipole moments of atomic spins. It is a simple, well studied and exactly solvable model from statistical physics. Those properties make it an ideal testing ground for the application of machine learning methods to physical systems. Its Hamiltonian is 
\begin{equation}
H(S)= -J\sum_{\left<i,j\right>_{nn}}s_{i}s_{j}+h\sum_{i}s_{i}\ .
\label{eq:Ising}
\end{equation}
In the following, we examine the ferromagnetic Ising model $J=1$ on the square lattice with vanishing external magnetic field $h=0$. $\left<i,j\right>_{nn}$ means summing over nearest neighbors, and $S=\left(s_{1},\ldots,s_{n}\right)$ denotes a spin configuration, where $s_{i}\in\left\{1,-1\right\}$. By defining the free energy at a given inverse temperature $\beta=(k_{\text{B}}T)^{-1}$, 
\begin{equation}
F(\beta)=-\beta^{-1}\text{log} \left(\sum_{i}e^{-\beta H(S_{i})}\right) \ ,
\end{equation}
thermodynamic quantities such as the expectation value of the energy,
\begin{equation}
\begin{aligned}
\left<E\right>&=-\frac{\partial \left(\beta F\right)}{\partial \beta}\ ,
\end{aligned}
\label{eq:energy}
\end{equation}
can be extracted. By the Ehrenfest classification the Ising model has a second order phase transition, since the specific heat $C_{V}=\partial \left<E\right>/\partial \beta$ diverges at $T_{\text{c}}=2/\left(k_{\text{B}}\text{log}\left(1+\sqrt{2}\right)\right)$ \cite{Onsager1944}. 

In Landau-Ginzburg theory phases are classified via the order parameter, in the Ising model this is the magnetization 
\begin{equation}
\left<M\right>=\left.\frac{\partial F}{\partial h}\right|_{h=0}\ ,
\label{eq:magnetization}
\end{equation}
which is zero in the paramagnetic phase and finite in the ferromagnetic phase. Its derivative with respect to the temperature diverges at the critical temperature.\\

\subsection{SU(2) Lattice Gauge Theory}
\noindent In this paper we examine SU(2) Yang-Mills theory, which shows confinement, one of the most distinct features of QCD. To this end we employ lattice gauge theory. It was  originally proposed by Wilson and Wegner and builds on the idea of discretizing the Euclidean path integral such that the lattice spacing $a$ is a natural cutoff scale. This discretization gives a strong analogy to statistical physics and allows for Monte Carlo simulations of gauge theories. SU(2) gauge theory on the lattice is parametrized by so-called link variables $U_{\mu}^{x}\in \text{SU}(2)$. Each lattice point $x$ attaches to one link variable per dimension $\mu$. In this work, we use four-dimensional $N_\tau \times N_{s}^{3}$ spacetime lattices with $N_{\tau}=2$ (temporal direction) and $N_{s} = 8$ (spatial volume). The link variables $U_{\mu}^{x}$ are parametrized by four real parameters, 
\begin{equation}
U_{\mu}^{x}=a_{\mu}^{x}\mathbb{1}+i\left(b_{\mu}^{x}\sigma_{1}+c_{\mu}^{x} \sigma_{2}+d_{\mu}^{x}\sigma_{3}\right)\ ,
\end{equation}
where $\sigma_{i}$ are the the Pauli matrices, the coefficients obey $(a_{\mu}^{x})^2+(b_{\mu}^{x})^2+(c_{\mu}^{x})^2 +(d_{\mu}^{x})^2=1$. The trace of $U_{\mu}^{x}$ is given by $2\,a_{\mu}^{x}$, since the Pauli matrices are traceless. Link variables are objects that live on the links between two neighbouring sites. They transform under gauge transformations via 
\begin{equation}
U_{\mu}^{x}\rightarrow \Omega^{x}U_{\mu}^{x}\left(\Omega^{x+\hat{\mu}}\right)^{\dagger} \ ,
\label{eq:gauge_trafo}
\end{equation}
where $\hat{\mu}$ is the unit vector in direction $\mu$ and $\dagger$ denotes the hermitian conjugate. This transformation property ensures gauge invariance of observables. A sample lattice configuration collects all link variables on the lattice
\begin{align}
S=(\{U_{\mu}^{x}|\, x\in N_\tau \times N_s\times N_s\times N_s,\, \mu\in \{\tau,x,y,z\}   \})\ .
\end{align}
From equation \eqref{eq:gauge_trafo}, it can be shown that closed loops over link variables are gauge invariant objects. The action we use in our simulations is the lattice version of the Yang-Mills action
\begin{equation}
\begin{aligned}
&S_{\text{Wilson}}[U] =\beta_{\text{latt}}\sum_{x}\sum_{\mu<\nu}\text{Re tr}\left(\mathbb{1}-U_{\mu\nu}^{x}\right)\ ,
\end{aligned}
\end{equation}
where $\beta_{\text{latt}}$ is the lattice coupling. Here  $U_{\mu\nu}^{x} = U_{\mu}^{x}U_{\nu}^{x+\hat{\mu}}U_{-\mu}^{x+\hat{\mu}+\hat{\nu}}U_{-\mu}^{x+\hat{\nu}}$ is the smallest possible closed rectangular loop. The order parameter for the deconfinement phase transition is the expectation value of the Polyakov loop. The Polyakov loop
\begin{align}
L(\vec{x})&=\text{tr}\left(\prod_{x_0=0}^{N_{\tau}-1}U_{\tau}^x\right)\stackrel{N_\tau=2}{=}\text{tr}\left(U_{\tau}^{0,\vec{x}}U_{\tau}^{1,\vec{x}}\right)\nonumber \\[2ex]
&=2\left(a^{0,\vec{x}}_{\tau}a^{1,\vec{x}}_{\tau}-b^{0,\vec{x}}_{\tau}b^{1,\vec{x}}_{\tau}-c^{0,\vec{x}}_{\tau}c^{1,\vec{x}}_{\tau}-d^{0,\vec{x}}_{\tau}d^{1,\vec{x}}_{\tau}\right) \ ,
\label{eq:pl}
\end{align}
is another gauge invariant quantity. It is the trace of a closed loop that winds around time direction using periodic boundary conditions. The expectation value of the Polyakov loop is zero in the confined phase and finite in the deconfined phase.  Another way to look at confinement is center symmetry. In the confined phase, SU(2) lattice gauge theory is symmetric under so-called center symmetry transformations. They are given by a multiplication of all temporal links at a given time slice by $z\in\left\{-1,1\right\}$. Under those transformations, the Polyakov loop transforms as $L\rightarrow zL$. This provides a strong analogy to the Ising model, since one can view individual Polyakov loops as spins corresponding to either value of $z$. In the deconfined phase, this symmetry is broken and Polyakov loops favor one of the values of $z$.
\\~\\
More details on the simulations can be found in Appendix \ref{app:MCS}.

%%%%%%%%%%%%%%%%%%%%%%%%%%%%%%%%%%%%%%%%%%%%%%%%%%%%%%%

\section{Machine Learning Pipeline}

\subsection{Artificial Neural Networks}
\noindent In this work we employ feed-forward artificial neural networks as a tool to distinguish between two classes in the context of supervised learning. Supervised learning is the field in machine learning where the algorithm learns to classify labeled training data. After being successfully trained, the algorithm is able to predict the label of unseen test samples with high accuracy. A feed-forward artificial neural network is a directed weighted graph consisting of layers, where only connections between neurons of neighboring layers are allowed. It has been shown that such an artificial neural network, with sufficiently many parameters, can approximate any continuous function \cite{Cybenko1989,Hornik1991}. We consider a neural network as an approximation of the decision function $D$. The decision function assigns to each sample S a probability $P\in [0,1]$ to be in class 1. The decision boundary is a hyperplane in the space of the parameters of sample configurations defined by $D(S)=0.5$, where the neural network is most unsure about the correct label. If there exists an explicit quantity $Q(S)$ which is learned by the neural network, and which is responsible for the distinction between phases, we expect that a change in the quantity $Q$ is always related to a change in the prediction probability. Hence $\nabla Q || \nabla D$ in the vicinity of the decision boundary. In our neural networks the output can be written as $D(S)=\text{sigmoid} (\xi(S))$, where $\text{sigmoid}(x) = 1/(1+\exp(-x))$ maps the latent prediction $\xi(S)$ to a probability. It follows that $\nabla Q || \nabla \xi$ and thus Q can be expressed as a linear function of $\xi$ in a linearized regime close to the decision boundary $\xi(S)=w \, Q(S)+b$. In the following sections we determine the latent prediction $\xi$ and hence the decision function $D$ as a simple function of $S$. Eventually, this allows us to explicitly formulate the quantity Q on which the neural network bases its classification.
%%%%%%%%%%%%%%%%%%%%%%%%%%%%%%%%
\begin{center}
\begin{figure}[htb!]
\includegraphics[width=0.5\textwidth]{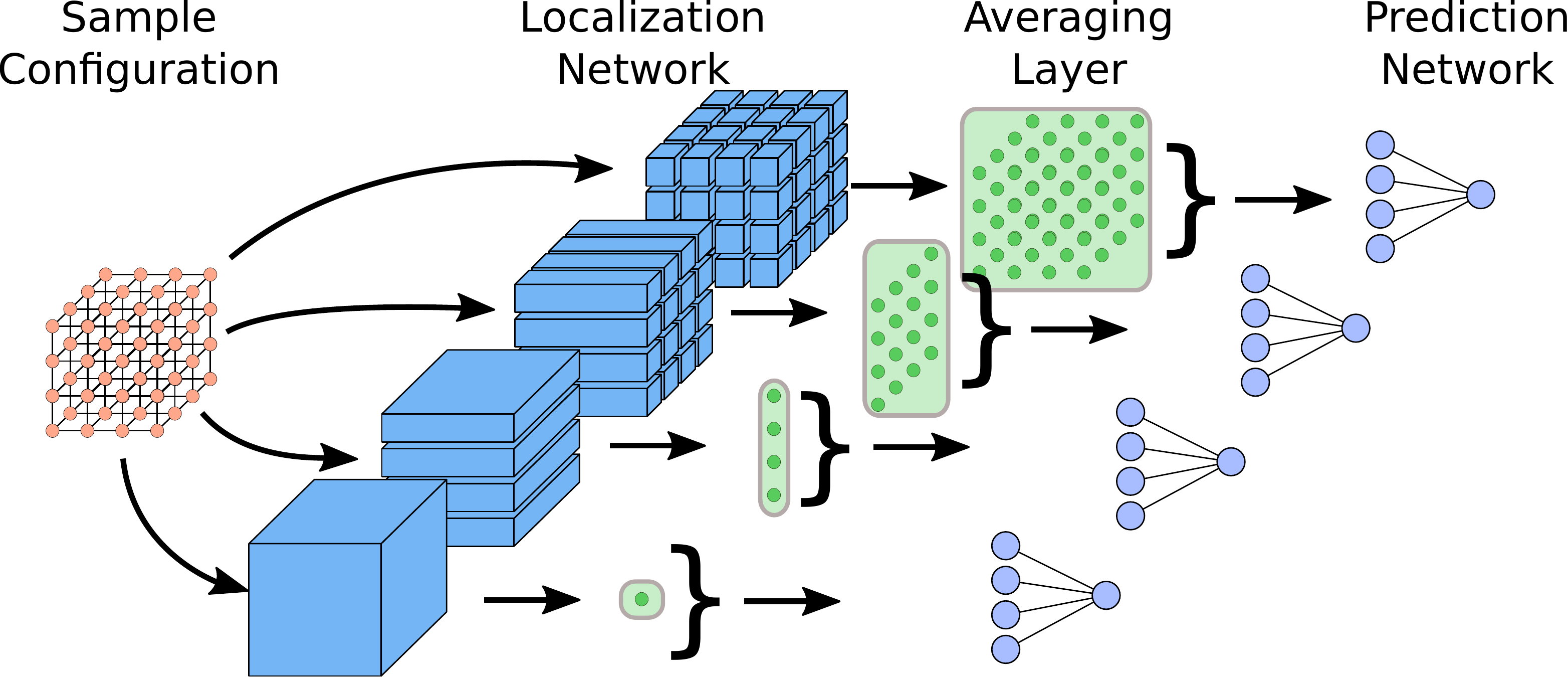}\\
\caption{The \textbf{Correlation Probing Neural Network} consists of three types of neural networks stacked on top of each other. The localization network is a fully convolutional neural network which prohibits connections outside of the receptive field of each output neuron. The averaging layer averages over the input from the localization network, similarly to how the magnetization averages over all spins. The prediction network is a fully connected neural network, which transforms the output of the averaging layer to a prediction probability.}
\label{fig:correlation_network}
\end{figure}
\end{center}
%%%%%%%%%%%%%%%%%%%%%%%%%%%%%%%%

\subsection{Ising Model}
\begin{figure*}[t!]
\includegraphics[width=0.9\textwidth]{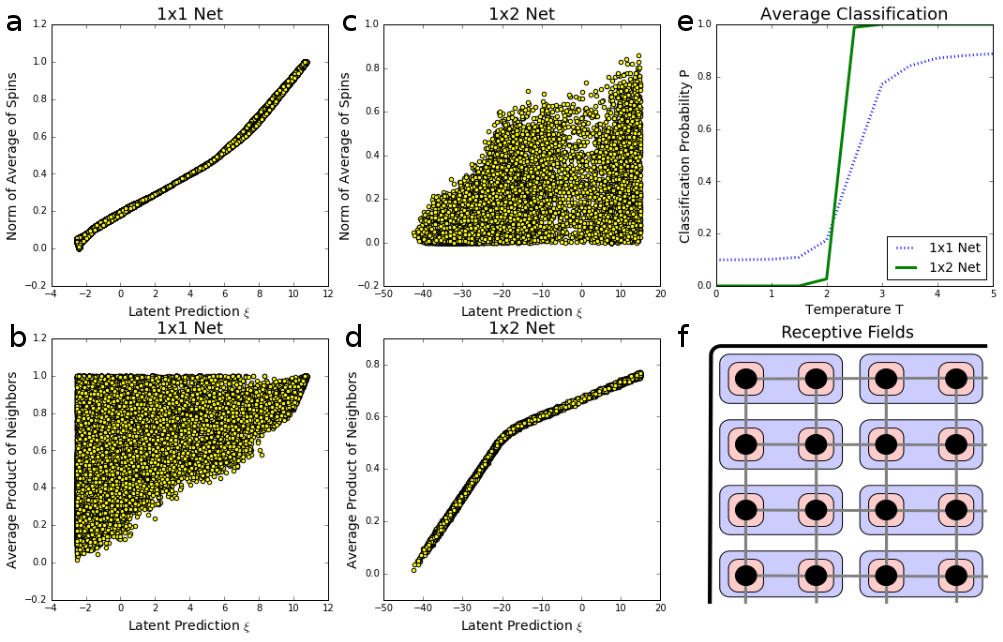}\\
\caption{Results of the correlation probing neural network applied to the Ising Model. The latent prediction $\xi$ is the argument of the $\text{sigmoid}$ function, which is the activation function in the last layer of the prediction network. \textbf{a ,b, c, d}: We calculated the values of either axis for all sample configurations and plotted them in scatterplots. \textbf{a}: The latent prediction is perfectly correlated with the absolute value of the average of spins, i.e the magnetization in the $1\times1$ network. \textbf{b}: It is not correlated with the average product of neighboring spins, i.e. the expected energy per site. \textbf{d}: The latent prediction of the $1\times2$ network is perfectly correlated with the average product of neighbors, but not with \textbf{c}: the norm of the average of spins. \textbf{e}: Average classification probability as a function of temperature. \textbf{f}: Receptive fields $1\times1$, $1\times2$ and $28\times28$ of the localization network output neurons, as probed in Table \ref{ising_losses}.}
\label{fig:ising_results}
\end{figure*}

\noindent Typically, order parameters are used to distinguish between phases. They are zero in one phase and finite in the other. For example in the Ising Model, see \sec{Sec:Ising}, the order parameter is the expectation value of the magnetization $M(S)=\left|1/N\sum_i s_i\right|$. Physical order parameters obey a lot of symmetries, thus they can be formulated in closed equations with few parameters. 

When machine learning algorithms and especially neural networks classify phases, they encode their decision function in a highly elusive and highly non-linear way. We pose the question if the decision function of a neural network can be expressed as a simple function of only a few specific spin correlations. In order to answer this question, we present a new type of neural network that is tailored to probe if specific correlations between spin variables contribute to the decision function of the neural network. We call it \emph{correlation probing neural network}, see \fig{fig:correlation_network}. The neural network architecture can be found in Appendix \ref{app:nnarchitecture}. It consists of three separate subnetworks: the localization network is a fully convolutional neural network which only allows connections between sites that have a predefined relative location to each other. In other words, this network consists of identical subnetworks acting only on patches of the input sample. The receptive field size of the output neurons of the localization network is the size of each of the patches. The output of the localization layer is averaged in the averaging layer, where all information about the spatial location is lost. The prediction network is a fully connected neural network which transforms the output of the averaging layer to a prediction. If, for example, the receptive field of a neuron of the localization layer has the dimensions $1\times1$, the correlation probing network can only approximate functions containing the correlation of a single spin variable with itself, as it is the case for the magnetization.

We apply the correlation probing neural network to the Ising Model by training it on Monte Carlo-sampled configurations below T=1.6 in the ordered phase and above T=2.9 in the unordered phase. More information about Monte Carlo simulation can be found in Appendix \ref{app:MCS}. The training objective is to correctly predict the phase of each sample configuration, which is achieved by minimizing the binary cross entropy loss function \eq{eq:crossentropy} between the correct label and the prediction. We compare the performance of the correlation probing neural network for different receptive field sizes in the localization network in Table \ref{ising_losses}. Using the full receptive field of $28\times28$, we allow the neural network to learn all possible spin correlations to approximate its decision function. In this case, the correlation probing network is equivalent to a standard convolutional neural network. The training and validation losses are minimized to a value close to zero. We conclude that the neural network has found all necessary information it needs to reliably classify the phases. By successively lowering the receptive field size, see \fig{fig:ising_results}\textbf{f}, we do not observe a drop in performance, except from $1\times2$ to $1\times1$ and from $1\times1$ to the baseline classifier, see Table \ref{ising_losses}. In each of these steps the neural network loses important information about the samples which it needs to reliably classify them. In \fig{fig:ising_results}\textbf{e} we can see the average classification probability, as a function of the temperature, of both networks. This plot, the training loss and the validation loss show that the $1\times1$ network is less accurate than the $1 \times 2$ network. The phase transition temperature can be found where $P=0.5$. This is at $T=2.5\pm 0.5$ for the $1\times1$ Network and $T=2.25\pm 0.25$ for the $1\times2$ Network. An accurate estimation can be found in \cite{Carrasquilla2017}. We however focus on examining what information got lost while lowering the receptive field size.

By construction, the decision function D of the $1\times 1$ neural network can be expressed as
\begin{align}
D(S)&=F\left(\frac{1}{N}\sum_i f(s_i)\right)\nonumber \\
 &=\text{sigmoid} \left(\xi \left( \frac{1}{N}\sum_i f(s_i)\right)\right)\ ,
\end{align}
where $F$ is the function approximated by the prediction network and $f$ is the function approximated by the localization network. The argument of $f$ is only a single spin. The function $f$ can be Taylor-expanded:
\begin{align}
f(s_i)=f_{0}+f_{1}\ s_i+f_{2} \ \underbrace{s_i^2}_{1}+ f_{3} \ \underbrace{s_i^3}_{s_i}+ ...
\end{align}
Since $s_i^2=1$, all higher order terms can be neglected. The constants $f_0$ and $f_1$ can be absorbed by the bias and the weights of the prediction network approximating $F$. Thus, the decision function reduces to
\begin{align}
D(S)=F\left(\frac{1}{N}\sum_i s_i\right) \ .
\end{align}
At this point we are almost done, since all $F$ does, is to formulate a probability $p\in[0,1]$ from its argument. In order to determine the function $F$, we need to compare the latent prediction $\xi$ of the neural network, with the argument of $F$: $1/N\sum_i s_i$, in the vicinity of the decision boundary. By looking at \fig{fig:ising_results}\textbf{a}, we infer that the latent prediction is given by $\xi(S)\approx w\left|1/N\sum_i s_i\right|+b$. This knowledge allows us to construct the function $F(x)\approx\text{sigmoid}(w\left| x\right|+b)$ and thereby the decision function 
\begin{align}
D(S)\approx\text{sigmoid}\left(w\ \left|\frac{1}{N}\sum_i s_i\right|+b\right) \ ,
\label{eq:dec1}
\end{align}
with weight $w$ and bias $b$ of the prediction neuron. 
The secondary purpose of this plot is to show the perfect correlation between the latent prediction $\xi(S)$ and $\left|1/N\sum_i s_i\right|$, which proves that our above deduction led to the correct result.

Until this point we have not used any information about the Ising model except Monte Carlo configurations. We have found that the decision function determines the phase by the quantity $Q(S)=\left|1/N \sum_i s_i \right|$. This function is the magnetization.

By examining the $1\times 2$, we require by construction that the decision function is of the form 
\begin{align}
D(S)=F\left(\frac{1}{N}\sum_{<i,j>_T} f(s_i,s_j)\right) \ .
\end{align}
Here the sum only goes over transversal nearest neighbors, collecting each spin only once. The Taylor expansion,
\begin{align}
f(s_i,s_j)=&f_{0,0}+f_{1,0}\, s_i+f_{0,1}\, s_j\nonumber \\
&+f_{2,0} \, s_i^2+ f_{1,1} \, s_i \, s_j+ f_{0,2} \, s_j^2+ ... \ ,
\end{align}
contains only three terms of note, all other terms can be reduced to simpler ones by using $s_i^2=1$. The terms $f_{1,0}\,s_i$ and $f_{0,1}\,s_j$ represent the magnetization. From Table \ref{ising_losses} and the analysis of the $1\times1$ network, we know that these terms contain less information than the quantity we are looking for. So the leading term must be $f_{1,1}s_is_j$. Thus, the decision function can be written as
\begin{align}
D(S)\approx F\left(\frac{1}{N}\sum_{<i,j>_T} s_i s_j\right)\ .
\end{align}
In \fig{fig:ising_results}\textbf{d} we see the perfect correlation between the latent prediction $\xi(S)$ and $1/N\sum_{<i,j>_T} s_i s_j$. This also means that the correction from the subleading terms $f_{1,0}\, s_i$ and $f_{0,1}\, s_j$ is indeed negligible. Hence we end up with the decision function 
\begin{align}
D(S)\approx\text{sigmoid}\left(w\left(\frac{1}{N}\sum_{<i,j>_T} s_i s_j\right)+b\right)\ .
\label{eq:dec2}
\end{align}
By translational and rotational symmetry, the sum can be generalized to all neighbors $Q(S)=\frac{1}{N}\sum_{<i,j>_{nn}} s_i s_j$. This quantity is, up to a minus sign, the expected energy per spin site. It is worth noting that the energy per site can be used to distinguish between phases more reliably than the magnetization, see Table \ref{ising_losses}. 

By examining the correlation probing neural network, we found which spatial correlations are crucial in deciding the phase. Furthermore, using this information, we reduced the complexity of the decision function so that we were able to fully reconstruct it. From the decision function one can read out the quantity by which the neural network makes its decision about phases. The quantities we found agree with the magnetization and the average energy per spin site. Both quantities are known to have diverging derivatives with respect to the temperature, and thus are defining quantities of a phase transition either in Landau-Ginzburg-theory or Ehrenfest classification.

\begin{table}[ht!]
\[
\begin{array}{c|c|c}
\hline
\text{Receptive Field Size} &\text{Train Loss}&\text{Validation Loss}\\
\hline
28\times28& 6.1588e-04& 0.0232\\
\hline
\textbf{1}\times\textbf{2}&\textbf{1.2559e-04} &\textbf{1.2105e-07}\\
\hline
\textbf{1}\times\textbf{1} &\textbf{0.2015}&\textbf{0.1886}\\
\hline
\text{baseline} &0.6931& 0.6931\\
\hline
\end{array}
\]
\caption{Ising model: Losses of neural networks with different receptive fields of the neurons in the localization network. This is a measure of how well a neural network performs, less is better. The baseline classifier is a random classifier which predicts each phase with a probability of $p=0.5$.}\label{ising_losses}
\end{table}

\subsection{SU(2) Lattice Gauge Theory}

\noindent We demonstrate the power of our new method at SU(2) lattice gauge theory in order to examine the deconfinement phase transition. We construct a whole machine learning pipeline around the correlation probing network, consisting of three different machine learning techniques: principal component analysis, neural networks and regression. Starting from no knowledge of an existing phase transition, we first employ unsupervised learning to get first indications about the existence and position of phases. Afterwards we train the correlation probing network to correctly predict phases. We then split the spin configurations to local configurations according to the spatial structure obtained from the correlation probing network. On these new samples, we train a local neural network to correctly classify phases. At last, we employ a regression algorithm to the local neural network results to find an explicit expression of the decision function of correlation probing neural network.

%%%%%%%%%%%%%%%%%%%%%%%%%%%%%%%%
\begin{center}
\begin{figure}[tb!]
\includegraphics[width=0.5\textwidth]{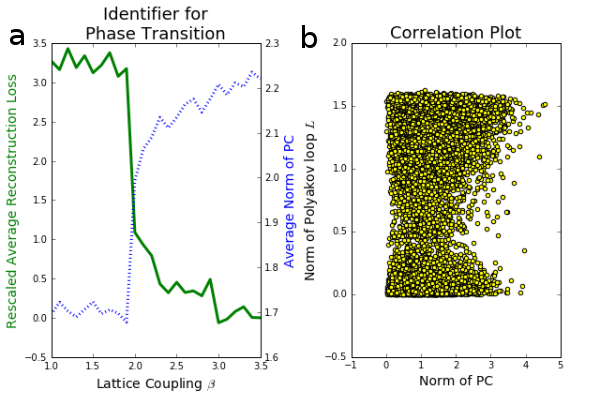}\\
\caption{\textbf{a}: Finding a possible phase transition with PCA. The average mean squared error reconstruction loss as a function of temperature is a universal identifier for a phase transition. It was calculated in 100 independent PCA runs with two principal components (PC), measured in units of $\times 10^{-5}$ and shifted by the value at $\beta=3.5$. The average norm of the PC also indicates a phase transition. \textbf{b}: There is no correlation between the principal components and the Polyakov loop.}
\label{fig:pca_results}
\end{figure}
\end{center}
%%%%%%%%%%%%%%%%%%%%%%%%%%%%%%%%

\subsubsection{Unsupervised Learning of Phase Transitions}

\noindent We assume no prior knowledge of the phase transition, even its existence. Hence, we employ unsupervised learning to find any possible indications for a phase transition. For the sake of simplicity we employ principal component analysis (PCA) \cite{Pearson1901,Wang2016} with two principal components. PCA is an orthogonal linear
transformation of the input samples to a set of variables, sorted by their variance. Here, unsupervised learning algorithms that are based on the reconstruction loss like autoencoders \cite{Wetzel2017} are doomed to fail, since the states are gauge invariant. The autoencoder would need to predict a matrix which is not unique.

Even though the Polyakov loop is a non-linear order parameter, PCA captures indications of a phase transition at $\beta \in [1.8,2.2]$, which is demonstrated in \fig{fig:pca_results}\textbf{a}. Here we employed the average reconstruction loss \cite{Wetzel2017} and the Euclidean norm of the principal components as identifiers for a phase transition. In \fig{fig:pca_results}\textbf{b} we show that there is no correlation between the Polyakov loop and the principal components.

It is worth noting that this example shows that PCA can capture phase indicators even when the principal components cannot approximate any order parameter.

\subsubsection{Correlation Probing Neural Network}
\begin{figure*}[t!]
\includegraphics[width=0.9\textwidth]{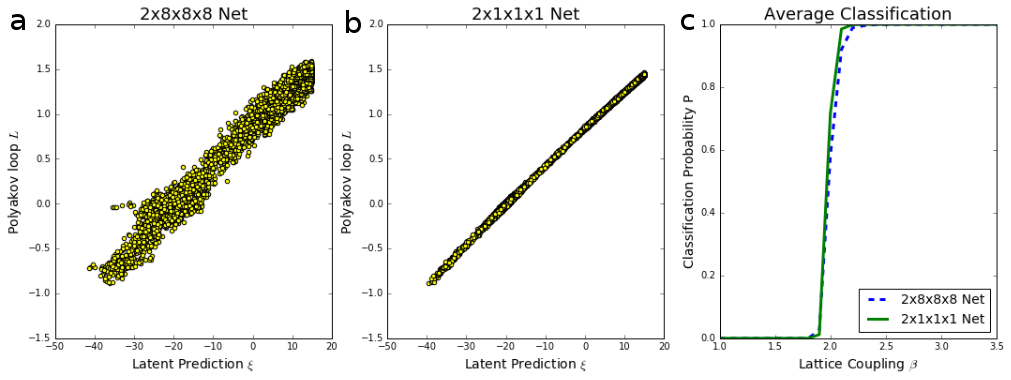}\\
\caption{Results of the correlation probing network applied to SU(2) lattice gauge theory. \textbf{a,b}: The latent prediction shows a strong correlation with the Polyakov loop in both the $2\times8\times8\times8$ network and the $2\times1\times1\times1$ network. \textbf{c}: The average prediction probability of the two networks. }
\label{fig:su2_results}
\end{figure*}
\noindent We train the correlation probing neural networks with different receptive fields, see \fig{fig:correlation_network} and Appendix \ref{app:nnarchitecture}, to predict phases on Monte Carlo-sampled configurations at lattice coupling $\beta \in[1,1.2]$ in one phase and $\beta\in[3.3,3.5]$ in the other phase. We test the neural network in $\beta \in[1.3,3.2]$ to predict a phase transition at $\beta = 1.99\pm0.10$ ($2\times1\times1\times1$ network) and $\beta=1.97\pm0.10$ ($2\times8\times8\times8$ network) , see \fig{fig:su2_results}\textbf{c}. A direct calculation from the lattice configurations reveals $\beta=1.880\pm0.025$, we comment on the difference in Appendix \ref{app:MCS}. By successively lowering the receptive field size we lose important information for classifying phases below $2\times1\times1\times1$, see Table \ref{su2_losses}. This means that crucial information about the phase transition is contained in this specific structure.

\begin{table}[t!]
\[
\begin{array}{c|c|c}
\hline
\text{Receptive Field Size} &\text{Train Loss}&\text{Validation Loss}\\
\hline
2\times8\times8\times8& 1.0004e-04& 2.6266e-04\\
\hline
2\times8\times1\times1& 8.8104e-08& 1.3486e-07\\
\hline
2\times1\times1\times1 &7.7671e-05& 2.0394e-04\\
\hline
\textbf{2}\times\textbf{1}\times\textbf{1}\times\textbf{1}&\textbf{8.8104e-08} &\textbf{ 6.8276e-08}\\
\hline
 \begin{tabular}{@{}c@{}}$\textbf{2}\times\textbf{1}\times\textbf{1}\times\textbf{1}$ \\ no hidden layers\\ in prediction net\end{tabular}&\textbf{2.2292e-07} &\textbf{ 4.2958e-07}\\
\hline
1\times1\times1\times1 &0.6620& 0.9482\\
\hline
\text{baseline} &0.6931& 0.6931\\
\hline
\end{array}
\]
\caption{SU(2):  Losses of neural networks with different receptive fields of the neurons in the localization network.}\label{su2_losses}
\end{table}

The decision function of the $2\times1\times1\times1$ network is constrained to
\begin{align}
D(S)=F\left( \frac{2}{N}\right.\sum_{\vec{x}}f(&U_\tau^{0,\vec{x}},U_x^{0,\vec{x}},U_y^{0,\vec{x}},U_z^{0,\vec{x}},\nonumber \\
&\left. \vphantom{\frac{2}{N}}  U_\tau^{1,\vec{x}},U_x^{1,\vec{x}},U_y^{1,\vec{x}},U_z^{1,\vec{x}})\right) \ ,
\end{align}
where $F$ is a function of a single variable and $f$ is a function of 32 variables, since each $U_\mu^x$ is uniquely defined by four real numbers. In order to determine the decision function, one could perform a combined polynomial fit of $F$ and $f$ on the latent prediction $\xi$. Since a feasible approach requires some knowledge about neural network architecture, we present this procedure in Appendix \ref{app:plregression}. \\

Here we present a different approach. It is based on reducing the expressibility of the neural network even further and separating the lattice to $2\times1\times1\times1$ patches. First, we convince ourselves that we do not need any hidden layers in the prediction network, i.e. we only keep the output neuron, see Table \ref{su2_nn}. Then the decision function simplifies to $D(S)= \text{sigmoid}(w\,Q(S)+b)$, where  
\begin{align}
Q(S)=\frac{2}{N}\sum_{\vec{x}}f(&U_\tau^{0,\vec{x}},U_x^{0,\vec{x}},U_y^{0,\vec{x}},U_z^{0,\vec{x}},\nonumber \\
&U_\tau^{1,\vec{x}},U_x^{1,\vec{x}},U_y^{1,\vec{x}},U_z^{1,\vec{x}}) \ 
\end{align}
reduces to a sum of functions acting only on a single patch of size $2\times1\times1\times1$ each. This allows us to split all samples to a minimum size of $2\times1\times1\times1$. By doing this, we enlarge the number of training samples by a factor of 512, which can enormously boost the accuracy in the following steps.
\subsubsection{Local Neural Network Regression}

\noindent We train a new neural network on the local data samples to classify the phases of each local sample. Although there is a lot more margin for error, the local neural network can now associate a prediction to each patch. 

We perform a polynomial regression on the latent prediction of the local neural network on only 1\% of the dataset and use another 1\% as validation set. We compare different orders of regression and find that a second order polynomial performs best, see Table \ref{regression_scores}. The regression approximates the latent prediction by a sum of 561 terms. We extract the weights of the regression and find that the parameter which quantifies the phase transition is given by
\begin{align}
f(\{U_{\mu}^{x_0}\})=&f(U_\tau^{0},U_x^{0},U_y^{0},U_z^{0},U_\tau^{1},U_x^{1},U_y^{1},U_z^{1})\nonumber \\
=&f((a_\tau^0,b_\tau^0,...,d_z^1))\nonumber \\
\approx&+7.3816\ a^{0}_{\tau} a^{1}_{\tau} +0.2529\ a^{1}_{\tau}b^{1}_{\tau}\nonumber \\
&+ \  ...\nonumber\\
&-0.2869\ d^{0}_{\tau}c^{1}_{\tau}-7.2279\ b^{0}_{\tau} b^{1}_{\tau}\nonumber \\
&-7.3005\ c^{0}_{\tau} c^{1}_{\tau}-7.4642\ d^{0}_{\tau} d^{1}_{\tau} \ .
\end{align}
We only keep the leading contributions and assume that the differences between the leading contributions originate from approximation errors. In this way we can justify errors of the size of the next to leading coefficients. Since overall factors and intercepts can be absorbed in the weights and biases of the neural network, we can simply rescale the above parameter to
\begin{align}
f((a_\tau^0,b_\tau^0,...,d_z^1))\approx&\, a^{0}_{\tau} a^{1}_{\tau}-b^{0}_{\tau} b^{1}_{\tau}-c^{0}_{\tau} c^{1}_{\tau}-d^{0}_{\tau} d^{1}_{\tau}\nonumber \\
=&\,  \tr \left(U_\tau^{0}U_\tau^{1}\right).
\end{align}
This is the Polyakov loop on a single spatial lattice site \eq{eq:pl}. We promote $f(\{U_{\mu}^{x_0}\})\rightarrow f(\{U_{\mu}^{x_0,\vec{x}}\})$ to act on the full lattice, such that we can formulate the decision function of the neural network with the full receptive field as 
\begin{align}
D(S)\approx\text{sigmoid}\left(w \left(\frac{2}{N}\sum_{\vec{x}} f(\{U_{\mu}^{x_0,\vec{x}}\})\right)+b\right) \ .
\label{eq:dec3}
\end{align}
Here $Q(S)=\frac{2}{N}\sum_{\vec{x}} f(\{U_{\mu}^{x_0,\vec{x}}\})$ is the Polyakov loop on the full lattice.
A confirmation of this deduction can be seen in the perfect correlation between the latent prediction and the Polyakov loop in \fig{fig:su2_results}\textbf{a,b}.
\begin{table}[htb!]
\[
\begin{array}{c|c|c}
\hline
\text{Order of Regression} &\text{Train Score}&\text{Validation Score}\\
\hline
\text{Polynomial Regression}&&\\
\hline
1&0.00128 &-0.00042\\
\hline
\textbf{2}&\textbf{0.72025}& \textbf{0.72395}\\
\hline
3&0.75675 & 0.69129\\
\hline
\hline
\text{Support Vector Regression}&&\\
\hline
1&-0.08943&-0.08988\\
\hline
2&0.64048&0.65367\\
\hline
3&-0.08434&-0.08963\\
\hline
\end{array}
\]
\caption{Scores of different regression algorithms. Higher is better.}\label{regression_scores}
\end{table}
\section{Conclusion}

\noindent We have combined the analyses of previous works and presented a pipeline of several machine learning algorithms which find the existence of different phases and predict the position of the phase transition. The most important result of our work is the explicit calculation of the decision functions of the neural networks classifying the phases of the Ising model, see equations \eq{eq:dec1} and \eq{eq:dec2}, and the decision function of the neural network applied to SU(2) lattice gauge theory, see equation \eq{eq:dec3}. 

For this purpose, we proposed the correlation probing neural network. By employing this network we analyzed the complexity and the symmetries of the decision function. The results of the correlation probing network enabled us to reconstruct the decision function in a simple form and thereby reveal the explicit formula of the quantity by which the neural network distinguishes between phases. The method was introduced at the Ising model on the square lattice, where the decision functions contain the formulas of the magnetization or of the expected energy per site. We then demonstrated the power of this new method at SU(2) lattice gauge theory, where the reconstructed decision function reveals the explicit mathematical expression of the Polyakov loop, a non-linear, non-local order parameter.

As of now machine learning is able to construct formulas of physical quantities relevant to phase transitions. This could find application in strongly correlated systems where the nature of the phases is unknown. Such systems include QCD at high densities and the two dimensional Hubbard model at finite chemical potential.

We hope to see further developments based on our approach: (i) It would be interesting to see how to extract order parameters when it is not possible to find a reduction to a semi local order parameter. This could be the case for example in the incommensurate antiferromagnetic phase in high-temperature superconductors. (ii) Machine learning performs better if it is trained on appropriate features. Our approach could also be used to decide on what features to generate from raw data, similar to, or as an extension to quantum loop tomography \cite{Zhang2016}. (iii) Our method of finding an order parameter can be expanded to unsupervised learning by embedding it into an autoencoder \cite{Wetzel2017}. (iv) It would also be interesting to study if this new approach can give insight in what neural networks learn in other disciplines, such as computer vision.

%%%%%%%%%%%%%%%%%%%%%%%%%%%%%%%%

\emph{Acknowledgments} We would like to thank Jan M. Pawlowski, Manfred Salmhofer, Ion-Olimpiu Stamatescu  and Christof Wetterich for useful discussions. We thank Shirin Nkongolo for reviewing the manuscript. S.W. acknowledges support by the Heidelberg Graduate School of Fundamental Physics. M.S. is supported by the DFG via project STA283/16-2.
%%%%%%%%%%%%%%%%%%%%%%%%%%%%%%%%

\appendix
\section{Monte Carlo Simulations}
\label{app:MCS}
\noindent In statistical physics and lattice gauge theory, Markov Chain Monte Carlo algorithms are used to sample lattice configurations from the Boltzmann factor. This is done by constructing a stochastic sequence that starts at some random initial configuration. This stochastic sequence is constructed such that the configurations obey Boltzmann statistics in the equilibrium. For more details on algorithm requirements and algorithms see e.g. \cite{Gattringer2010}.

Observables are then computed by taking the average over many spin or lattice configurations from the equilibrium distribution
\begin{equation}
\left<\mathcal{O}\right>= \lim\limits_{N \to \infty} \frac{1}{N}\sum_{i=1}^{N}\mathcal{O}_{i} \ .
\end{equation}
Taking the limit in the last equality is practically not possible. Hence, the expectation value of the observable is approximated by large $N$ and gives rise to a statistical error. It is important to take enough configurations such that ergodicity is achieved. In the case of two distinct regions of phase space, this can take very long simulation time.

For the Ising model, we produced a total of 55000 spin configurations, of size $28\times 28$, equally distributed over eleven equidistant temperature values 
 $T \in\left[0,5\right]$ by employing the Metropolis-Hastings algorithm \cite{Metropolis1949} with simulated annealing.
 
For SU(2), we used the Heatbath algorithm \cite{Creutz1980} to produce a total of 15600 decorrelated configurations  equally distributed over 26 values in the range of $\beta_{\text{latt}}=4/g^{2}\in\left[1,3.5\right]$. In the context of this paper it is important to have decorrelated data, since neural networks are good at finding structures, and thus correlations between configurations, if existent. Due to center symmetry breaking, in the deconfined phase the average Polyakov loop can take either positive or negative values of equal magnitude, hence one usually takes the absolute value as an order parameter. At large values of $\beta_{\text{latt}}$, this will prevent a full exploration of phase space. In our simulations, we initiated all links with the unit matrix, hence we introduced a bias for large values of $\beta_{\text{latt}}$, i.e. our simulations are not fully ergodic.
If we were to employ neural networks  to extract the position of the phase transition, this non-ergodicity would lead to a shift in the value of critical $\beta_{\text{latt}}$. Generally speaking, ergodicity can be retained by doing more simulations and employing algorithms such as simulated annealing or overrelaxation, thus in principle it should be possible to extract the critical temperature reliably.

%%%%%%%%%%%%%%%%%%%%

\section{Neural Network Architecture}
\label{app:nnarchitecture}

\noindent We constructed our machine learning pipeline using Scikit-learn \cite{pedregosa2011} and Keras \cite{chollet2015}. The neural network architectures are presented in Tables \ref{ising_nn} and \ref{su2_nn}. Since there is no Convolutional4D in Keras, we just rearranged our samples to fit a Convolutional1D Layer. We used neural networks with number of filters $n_A,n_B,n_D \in \{1,4,8,32,256,1024\}$. The kernel sizes A, B, C are used to set the receptive field size. For our problems, $n_C=1$ is sufficient to capture the structure of the order parameter. This was probed in the same manner as finding the optimal receptive field size. In other models one might need a higher $n_C$, e.g. in the Heisenberg model, $n_C=3$ could be optimal. Hence, this can already be an early indicator for the type of the broken symmetry. The activation functions are rectified linear units $\text{relu}(x)=\max(0,x)$ between all layers and the sigmoid function $\text{sigmoid}(x)=1/(1+\exp(-x))$ in the last layer. We do not employ any sort of regularization. The training objective is minimizing the binary cross entropy loss function
\begin{align}
C(Y,P)=-\frac{1}{N}\sum_i(y_i \log p_i+ (1-y)\log(1-p_i))\ ,
\label{eq:crossentropy}
\end{align}
where $Y={y_i}$ is a list of labels and $P={p_i}$ is the corresponding list of predictions. Our baseline classifier is the classifier which assigns each label with a probability of $p_i=0.5$. This means that this classifier just assigns a label to each sample randomly. The binary cross entropy then evaluates to $0.6931$. The neural networks learn by optimizing the weights and biases via RMSprop gradient descent. The neural networks were trained for 300 epochs or less, if the loss already saturated in earlier epochs. The validation set is $20\%$ of the training data.
\begin{table}[ht!]
\[
\begin{array}{c|c|c}
\hline
\text{Layer}          &          \text{Output Shape}     & \text{Kernel Size}  \\    
\hline

InputLayer     &        (784, 1) &  \\

Convolution1D & (784/(A), n_A) & A \\

Convolution1D & (784/(A\times B), n_B) &B \\

Convolution1D&  (784/(A\times B\times C),n_C) &C\\

Average Pooling& (1, n_C)   &\\

Flatten     &         (n_C)   & \\

Dense       &           (n_D)&  \\

Dense        &          (1) &     \\
\hline
\end{array}
\]
\caption{Ising Model Neural Network. A, B, C determine the receptive field size of each neuron in the averaging layer.}\label{ising_nn}
\end{table}

\begin{table}[ht!]
\[
\begin{array}{c|c|c}
\hline
\text{Layer}                &    \text{Output Shape}     &     \text{Kernel Size}     \\       
\hline
InputLayer       &      ( 1024, 16)     &                                  \\        

Convolution1D & (1024/(A), n_A)  &     A          \\

Convolution1D&  ( 1024/(A\times B), n_B)&      B \\
Convolution1D & ( 1024/(A\times B \times C), n_C)&      C    \\

Average Pooling& ( 1, n_C) &       \\

Flatten       &      ( n_C)          & \\

Dense         &         (n_D)          &             \\

Dense      &           ( 1)         &     \\
\hline
\end{array}
\]
\caption{SU(2) Neural Network. A, B, C determine the receptive field size of each neuron in the averaging layer.}\label{su2_nn}
\end{table}

%%%%%%%%%%%%%%%%%%%%%%%%%%%%%%

\section{Regression of the Polyakov Loop}
\label{app:plregression}
\noindent The decision function of the $2\times1\times1\times1$ neural network which predicts the lattice SU(2) phase transition is by construction
\begin{align}
D(S)=F\left( \frac{2}{N}\right.\sum_{\vec{x}}f(&U_\tau^{0,\vec{x}},U_x^{0,\vec{x}},U_y^{0,\vec{x}},U_z^{0,\vec{x}},\nonumber \\
&\left. \vphantom{\frac{2}{N}}  U_\tau^{1,\vec{x}},U_x^{1,\vec{x}},U_y^{1,\vec{x}},U_z^{1,\vec{x}})\right) \ .
\end{align}

In general, we cannot assume that the prediction network consists only of the output neuron. Therefore, we suggest a different procedure for constructing the decision function. We split the full correlation probing net into subnetworks: we extract the localization network plus averaging layer and the prediction network as separate networks. In order to determine $F(S)=\text{sigmoid}(\xi(S))$, we use polynomial regression to fit the latent prediction of the prediction network to the output of the averaging layer. We find a polynomial of degree 1 is enough to fit the data, and $\xi$ is approximated by
\begin{align}
\xi(x)\approx-0.7101 \, x+ 9.85143419 \ .
\end{align}
The slope and intercept can be absorbed by the weight $w$ and bias $b$ of the output neuron, such that we can infer
\begin{align}
\xi(x)\approx w\, x +b\ .
\label{eq:app1}
\end{align}
 The function $f$ requires us to build a new local neural network which only acts on patches of size $2\times1\times1\times1$. By construction this network has the same number of weights and biases, as the full neural network acting on the input of size $2\times8\times8\times8$. Instead of training the local neural network, we transfer the the weights and biases from the full correlation probing network to the local neural network. Hence, one can obtain the output of the localization network for each patch separately. Again, we employ polynomial regression to fit the input from the local patches to the output of the localization network. The result of a regression of degree 2 with 561 parameters yields
\begin{align}
f(\{U_{\mu}^{x_0}\})=&f(U_\tau^{0},U_x^{0},U_y^{0},U_z^{0},U_\tau^{1},U_x^{1},U_y^{1},U_z^{1})\nonumber \\
=&f((a_\tau^0,b_\tau^0,...,d_z^1))\nonumber \\
\approx&-26.8354\ a^{0}_{\tau} a^{1}_{\tau} -2.4972\ d^{0}_{\tau}c^{1}_{\tau}\nonumber \\
&+ \  ...\nonumber\\
&+1.5653\ b^{0}_{\tau}c^{0}_{\tau}+26.5908\ b^{0}_{\tau} b^{1}_{\tau}\nonumber \\
&+27.7054\ c^{0}_{\tau} c^{1}_{\tau}+27.8939\ d^{0}_{\tau} d^{1}_{\tau} \ .
\end{align}
After absorbing overall factors and the intercept by the weights and biases of the prediction network and neglecting the subleading terms, we rewrite $f$ as
\begin{align}
f((a_\tau^0,b_\tau^0,...,d_z^1))\approx a^{0}_{\tau} a^{1}_{\tau} -\ b^{0}_{\tau} b^{1}_{\tau}- c^{0}_{\tau} c^{1}_{\tau}- d^{0}_{\tau} d^{1}_{\tau} \ .
\label{eq:app2}
\end{align}
This is the Polyakov loop on a single lattice site \eq{eq:pl}. 
By employing \eq{eq:app2} as an argument of \eq{eq:app1}, we can promote $f(\{U_{\mu}^{x_0}\})\rightarrow f(\{U_{\mu}^{x_0,\vec{x}}\})$ to depend on space again. We obtain the definition of the decision function
\begin{align}
D(S)\approx\text{sigmoid}\left(w \left(\frac{2}{N}\sum_{\vec{x}} f(\{U_{\mu}^{x_0,\vec{x}}\})\right)+b\right) \ ,
\end{align}
where $Q(S)=\left(\frac{2}{N}\sum_{\vec{x}} f(\{U_{\mu}^{x_0,\vec{x}}\})\right)$ is the Polyakov loop on the full lattice.

\bibliographystyle{unsrtnat}

\bibliography{VAEIsingModelBib}

%%%%%%%%%%%%%%%%%%%%%%%%%%%%%%%%

\end{document}